\documentclass[twocolumn,showpacs,preprintnumbers]{revtex4}
\usepackage{amssymb}
\usepackage{amsmath}
\usepackage[english]{babel}
\usepackage[dvips]{graphicx}

\newcommand{\be}{\begin{equation}}
\newcommand{\ee}{\end{equation}}
\newcommand{\bea}{\begin{eqnarray}}
\newcommand{\eea}{\end{eqnarray}}

\begin{document}

\title{Solitons in a chain of $\mathcal{PT}$-invariant dimers}

\author{Sergey V. Suchkov$^{1}$, Boris A. Malomed$^{2,3}$, Sergey V. Dmitriev%
$^{1}$, and Yuri S. Kivshar$^{4}$}
\affiliation{$^1$Institute for Metals Superplasticity Problems, Russian Academy of
Science, Ufa 450001, Russia \\
$^{2}$Department of Physical Electronics, School of Electrical Engineering,
Faculty of Engineering, Tel Aviv University, Tel Aviv 69978, Israel \\
$^{3}$ICFO-Institut de Ciencies Fotoniques, Mediterranean Technology Park,
Castelldefels 08860, Spain\thanks{%
Sabbatical address}\\
$^{4}$Nonlinear Physics Center, Research School of Physics and Engineering,
Australian National University, Canberra ACT 0200, Australia}

\begin{abstract}
Dynamics of a chain of interacting parity-time invariant nonlinear dimers is
investigated. A dimer is built as a pair of coupled elements with equal gain
and loss. A relation between stationary soliton solutions of the model and
solitons of the discrete nonlinear Schr\"{o}dinger (DNLS)\ equation is
demonstrated. Approximate solutions for solitons whose width is large in
comparison to the lattice spacing are derived, using a continuum counterpart
of the discrete equations. These solitons are mobile, featuring nearly
elastic collisions. Stationary solutions for narrow solitons, which are
immobile due to the pinning by the effective Peierls-Nabarro potential, are
constructed numerically, starting from the anti-continuum limit. The
solitons with the amplitude exceeding a certain critical value suffer an
instability leading to blowup, which is a specific feature of the nonlinear $%
\mathcal{PT}$-symmetric chain, making it dynamically different from DNLS
lattices. A qualitative explanation of this feature is proposed. The
instability threshold drops with the increase of the gain-loss coefficient,
but it does not depend on the lattice coupling constant, nor on the
soliton's velocity.
\end{abstract}

\maketitle

\section{Introduction}

In the original works by C. Bender and co-authors \cite{Bender,Bender-review}%
, it was pointed out that non-Hermitian Hamiltonians can have an entirely
real eigenvalue spectrum under the parity-time ($\mathcal{PT}$) symmetry
constraint. This mathematical observation can have deep physical
consequences, essentially altering the familiar properties of collective
modes in the respective media combining amplification and dissipation.
Indeed, it is usually assumed that the balance between the gain and loss
uniquely selects parameters of isolated stable modes. However, the use of
the $\mathcal{PT}$ symmetry makes it possible to support continuous families
of modes, allowing the dissipative media to emulate conservative ones, up to
a certain threshold level. In particular, the effects of amplification and
dissipation may stay in the exact balance for symmetric configurations of
weak fields, hence the $\mathcal{PT}$-invariant dynamics is preserved. In
contrast, for field intensities above the threshold, nonlinear self-action
(if present in the system) breaks the $\mathcal{PT}$ symmetry both locally
and globally, resulting in the asymmetric wave localization in the region
with amplification.

As the necessary condition for the $\mathcal{PT}$-symmetry of the
Hamiltonian with a complex potential, $V(x)$, is reduced to condition $%
V(x)=V^{\ast }(-x)$, such systems can be realized in the most
straightforward way in optics, by combining a spatially symmetric profile of
the refractive index with symmetrically placed mutually balanced gain and
loss~\cite{Ruschhaupt} (complex potentials may also be physically relevant
in the case when they are not subject to the $\mathcal{PT}$-symmetry
constraint \cite{VVK}). The possibility to realize physical systems with the
$\mathcal{PT}$ symmetry was a motivation for many theoretical~\cite{theory}
and experimental~\cite{experiment} works. Effects of the conservative
nonlinearity in $\mathcal{PT}$-symmetric systems were addressed too \cite%
{nonlinear,Coupler,OL}. Recently, the nonlinearity of the gain and loss,
also subject to the condition of the $\mathcal{PT}$ symmetry, was introduced
in Ref. \cite{nonlin2}.

The simplest nonlinear $\mathcal{PT}$-symmetric object can be realized as a
pair of linearly-coupled optical waveguides (alias a \textit{dimer}),
composed of a passive waveguide carrying linear loss and its active
counterpart imparted with a matched compensating gain \cite{Coupler} (in a
more general form, without the condition of the exact equilibrium between
the loss and gain, the same system of linearly coupled active and passive
waveguiding cores was investigated for a long time in various forms \cite%
{Winful}-\cite{Dmitry}, see also a review in Ref. \cite{Chaos}). A chain of
such $\mathcal{PT}$-invariant couplers was studied in Ref. \cite{OL}, with
each active or passive element linearly connected with an element of the
opposite sign, belonging to an adjacent coupler. In other words, in this
setting the axis of each waveguiding coupler is \emph{perpendicular} to the
direction along which the chain is built. It was demonstrated that this
chain of couplers can support stable solitons with amplitudes smaller than a
threshold amplitude.

In the present work, we propose and analyze another array of $\mathcal{PT}$%
-symmetric couplers with the intrinsic nonlinearity. In the array, active
and passive elements are linearly coupled to the elements of the same sign
belonging to adjacent dimers [see Fig. \ref{Fig1}(a)], i.e., the axes of the
couplers are aligned with the direction of the chain composed of them. In
fact, the proposed system is quite natural in terms of the realization in
optics, as it can be built of two parallel extended waveguides, one pumped
and one lossy, if each one is segmented into an array of individual
waveguides. Our aim is to find nonlinear localized modes (discrete solitons)
in this chain, and study their stability, mobility, and interaction.

The paper is organized as follows. Section \ref{Model} outlines the model.
Analytical results for moving and stationary solitons and their stability
are presented in Sec. \ref{SolitonSolutions}. Results of numerical studies
are reported in Sec. \ref{NumericalResults}, and Sec. \ref{Conclusions}
concludes the paper.

\section{Model}

\label{Model}

\begin{figure}[tbp]
\includegraphics[width=\columnwidth]{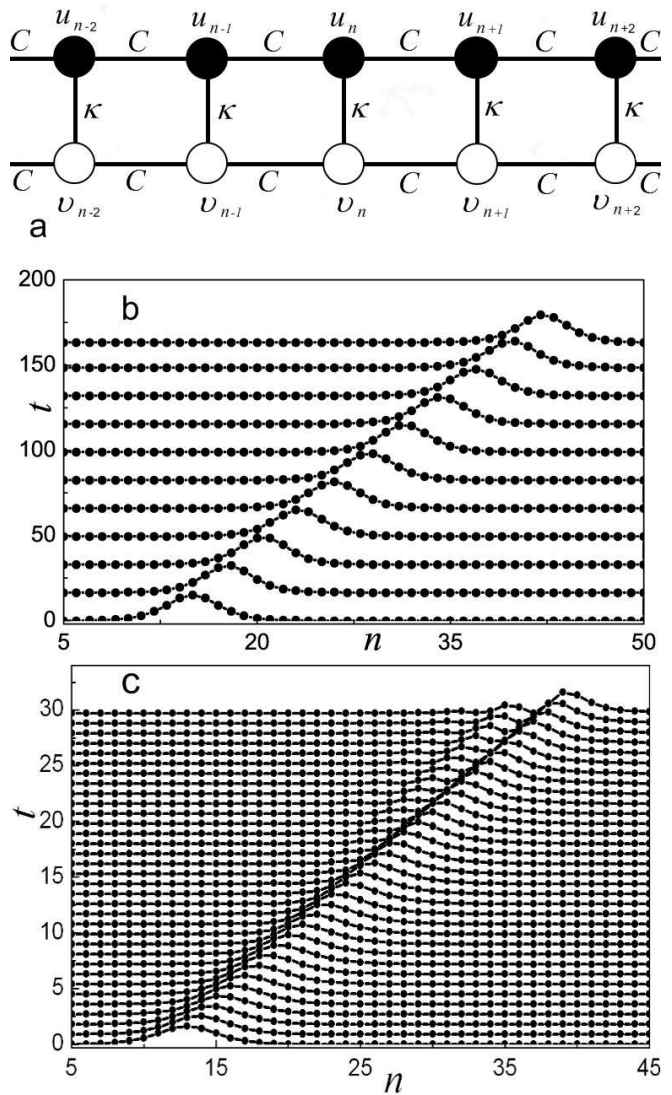}
\caption{(a) A schematic of the chain of dimers. (b) The spatio-temporal
evolution of $|u_{n}(t)|^{2}$ showing the propagation of a soliton through
the chain of dimers described by Eq. (\protect\ref{vnorm}) [the respective
plot for $|v_{n}(t)|^{2}$ is nearly identical to this one]. The initial
conditions were set as per the approximate solution (\protect\ref{discsol}),
with $\protect\gamma =0.5$, $A=0.37$, $C=0.5$, $V=0.2$. (c) Same as in (b),
but for a larger velocity, $V=2$.}
\label{Fig1}
\end{figure}

As said above, we consider a nonlinear chain shown in Fig.~\ref{Fig1}(a),
which is described by the following set of equations:
\begin{eqnarray}
\frac{du_{n}}{dt} &=&+\gamma u_{n}+i\sigma |u_{n}|^{2}u_{n}+i\kappa v_{n}+
\notag \\
&&iC\left( u_{n+1}+u_{n-1}-2u_{n}\right) ,  \notag \\
\frac{dv_{n}}{dt} &=&-\gamma v_{n}+i\sigma |v_{n}|^{2}v_{n}+i\kappa u_{n}+
\notag \\
&&iC\left( v_{n+1}+v_{n-1}-2v_{n}\right) .  \label{v1}
\end{eqnarray}%
Here $\gamma >0$ is the coefficient of the gain and loss acting on complex
variables $u_{n}$ and $v_{n}$, which correspond, respectively, to the active
and passive elements, real coefficient $\sigma $ accounts for the nonlinear
frequency shift (the intrinsic nonlinearity of each element), $\kappa $ is
the constant of the vertical coupling inside the dimer, and $C$ is a
coefficient of the horizontal linear coupling between dimers in the chain,
see Fig. \ref{Fig1}(a). Using obvious transformations ($v_{n}\rightarrow
-v_{n}$, the complex conjugation, and staggering, respectively), one can fix
$\kappa $, $C$, and $\sigma $ to be positive. Finally, by means of
rescaling, we can set $\kappa =\sigma \equiv 1$, which leaves $\gamma >0$
and $C>0$ as irreducible parameters, the final form of the equations being
\begin{eqnarray}
\frac{du_{n}}{dt} &=&\gamma
u_{n}+i|u_{n}|^{2}u_{n}+iv_{n}+iC(u_{n+1}+u_{n-1}-2u_{n}),  \notag
\label{vnorm} \\
\frac{dv_{n}}{dt} &=&-\gamma
v_{n}+i|v_{n}|^{2}v_{n}+iu_{n}+iC(v_{n+1}+v_{n-1}-2v_{n}).  \notag \\
&&
\end{eqnarray}

First, we will derive approximate solutions for moving broad solitons, whose
width is large in comparison with the lattice spacing, and study their
almost elastic collisions, using a continuum counterpart of the discrete
equations. We will also consider the case of narrow solitons, finding such
strongly localized solutions numerically, starting from the anti-continuum
limit. It will be demonstrated that the solitons with the amplitude
exceeding a certain critical value become unstable. Because the instability
is induced by the interplay of the gain and loss in the $\mathcal{PT}$%
-symmetric system, the instability threshold naturally lowers with the
increase of the strength of the gain and loss.

\section{Solitons}

\label{SolitonSolutions}

\subsection{Analytical approximations}

We start with the analysis of the continuum limit ($C\rightarrow \infty $)
of Eq. (\ref{vnorm}), which leads to the system of PDEs,
\begin{eqnarray}
\frac{\partial u}{\partial t} &=&+\gamma u+i|u|^{2}u+iv+i\frac{\partial ^{2}u%
}{\partial x^{2}},  \notag \\
\frac{\partial v}{\partial t} &=&-\gamma v+i|v|^{2}v+iu+i\frac{\partial ^{2}v%
}{\partial x^{2}},  \label{vcont}
\end{eqnarray}%
which possess exact soliton solutions,%
\begin{equation}
\left\{
\begin{array}{c}
u(t,x) \\
v(t,x)%
\end{array}%
\right\} =A~\mathrm{sech}\left[ \frac{A}{\sqrt{2}}(x-Vt)\right] e^{i\left(
-\omega t+Vx/2\right) }\left\{
\begin{array}{c}
e^{i\delta /2} \\
e^{-i\delta /2}%
\end{array}%
\right\} .  \label{contsol}
\end{equation}%
Here $A$, $A^{-1}$, $V$, $\omega $, and $\delta $ represent the amplitude,
width, velocity, frequency, and intrinsic phase shift of the soliton:
\begin{eqnarray}
\omega  &=&-A^{2}-\cos \delta +V^{2}/4+A^{2}/2,~\sin \delta =-\gamma ,
\notag \\
\cos \delta  &=&\pm \Gamma ,\Gamma \equiv \sqrt{1-\gamma ^{2}}.
\label{parameters}
\end{eqnarray}%
In the following the case $\cos \delta =\Gamma $ will be considered. Note
that $\Gamma \leq 1$, and this solution exists for $\gamma \leq 1$. In the
limit of the vanishing gain and loss, $\gamma \rightarrow 1$, Eqs. (\ref%
{vcont}) become tantamount to the well-known model of dual-core nonlinear
optical fibers. In that case, soliton (\ref{contsol}) with $\cos \delta
=\Gamma \equiv +1$ reduces to the symmetric soliton in the dual-core fiber,
which loses its stability through the symmetry-breaking bifurcation at a
finite value of the energy \cite{dual}. For the comparison with the discrete
system introduced in the present work, it is important to mention that a
similar symmetry-breaking instability of two-component solitons in the
discrete counterpart of the dual-core-fiber model (in fact, this discrete
system may also be readily implemented in optics) was found in Ref. \cite%
{discrete}.

Although equations (\ref{vcont}) are dissipative, they are Galilean
invariant, hence the soliton may move at an arbitrary velocity, $V$. It is
also worthy to note that the amplitude of the soliton may be arbitrary,
i.e., the present solutions form a \emph{continuous family}, unlike formally
similar approximate \cite{Winful,Athens,Dmitry} and exact \cite{S1} soliton
solutions previously found in the above-mentioned models of dual-core
systems with the gain applied in one core, and loss acting in the other.
Those models were considered without the condition of the balance between
the gain and loss in the general case (although the case of equal gain and
loss had a special purport in that context too), therefore they support
solitons with two discrete values of the amplitude, one stable and one
unstable.

For sufficiently wide solitons, i.e., for $A\ll \sqrt{C}$, the following
\emph{approximate} solution to the discrete equation (\ref{vnorm}) can be
deduced from the continuum solution (\ref{contsol}),
\begin{gather}
\left\{
\begin{array}{c}
u_{n}(t) \\
v_{n}(t)%
\end{array}%
\right\} =A~\mathrm{sech}\left[ \frac{A}{\sqrt{2}}\left( \frac{n-x_{0}}{%
\sqrt{C}}-Vt\right) \right] \left\{
\begin{array}{c}
e^{i\delta /2} \\
e^{-i\delta /2}%
\end{array}%
\right\}  \notag \\
\times \exp \left\{ i\left[ \left( \frac{A^{2}}{2}+\Gamma -\frac{V^{2}}{4}%
\right) t+\frac{V(n-x_{0})}{2\sqrt{C}}\right] \right\} ,  \label{discsol}
\end{gather}%
where $x_{0}$ is shift along the lattice. An example of the soliton
propagation, generated by simulations of Eq. (\ref{vnorm}) with the initial
conditions taken as per approximation (\ref{discsol}), is displayed in Fig. %
\ref{Fig1}(b). In this case, $A/\sqrt{2C}=0.37$, i.e., the soliton is not
very wide; nevertheless, approximation (\ref{discsol}) yields good initial
conditions for the soliton. The moving soliton practically does not radiate
energy in the form of small-amplitude waves. On the contrary, Fig. \ref{Fig1}%
(c) demonstrates that the soliton with a large velocity lose energy through
the emission of radiation.

For narrow solitons, with $A\gtrsim \sqrt{C}$, the approximation based on
the continuum limit is obviously irrelevant. In this case, solitons can be
constructed numerically, starting from solutions which are exact ones in the
\textit{anticontinuum} (AC) \textit{limit}, that corresponds to $C=0$ in Eq.
(\ref{vnorm}):
\begin{eqnarray}
\left\{
\begin{array}{c}
u_{0} \\
v_{0}%
\end{array}%
\right\} &=&A\left\{
\begin{array}{c}
e^{i\delta /2} \\
e^{-i\delta /2}%
\end{array}%
\right\} e^{-i\omega t},  \notag \\
u_{n} &=&v_{n}=0\quad \mathrm{for}\quad n\neq 0,  \label{2column}
\end{eqnarray}%
where $A$ is an arbitrary amplitude, while the frequency and phase shift
between the two components are given by
\begin{equation}
\omega =-A^{2}-\cos \delta ,\quad \sin \delta =-\gamma ,\quad \cos \delta
=\pm \Gamma .  \label{A}
\end{equation}%
In the following the case $\cos \delta =\Gamma $ will be considered,
following the choice adopted above for the broad solitons.

It follows from Eq. (\ref{A}) that this solution does not exist at
\begin{equation}
\gamma >1~~\mathrm{or}~~\omega <-\Gamma .  \label{nothing}
\end{equation}%
On the other hand, for $C\neq 0$, one has the continuous-wave (CW) solution
to Eq. (\ref{vnorm}), i.e., the wave with a constant amplitude, in the form
of
\begin{equation}
u_{n}=u_{0},v_{n}=v_{0}  \label{CW}
\end{equation}%
for all $n$, where $u_{0}$ and $v_{0}$ are defined by Eq. (\ref{2column})
and Eq. (\ref{A}). %\begin{equation}
%\left\{
%\begin{array}{c}
%u_{n} \\
%v_{n}%
%\end{array}%
%\right\} =A\left\{
%\begin{array}{c}
%e^{i\delta /2} \\
%e^{-i\delta /2}%
%\end{array}%
%\right\} e^{-i\omega t},  \label{CW}
%\end{equation}%
%with parameters defined by

We stress that, as well as the exact soliton solutions (\ref{contsol}) and
approximate ones (\ref{discsol}), the AC and CW solutions, given by Eqs. (%
\ref{2column}) and (\ref{CW}), respectively, form continuous families in
spite of the fact that they exist in the dissipative system. This is a
fundamental manifestation of the $\mathcal{PT}$ symmetry of the present
system.

\subsection{Stability analysis}

The next step of the analysis is the consideration of the stability of the
AC solution (\ref{2column}) and CW solution given by Eq. (\ref{CW}). In the
latter case, the perturbed CW state is looked for as%
\begin{equation}
\left\{
\begin{array}{c}
\tilde{u}_{n} \\
\tilde{v}_{n}%
\end{array}%
\right\} =\left\{
\begin{array}{c}
\left( A+\epsilon _{1}e^{\alpha t-ikn}\right) e^{i\delta /2} \\
\left( A+\epsilon _{2}e^{\alpha t-ikn}\right) e^{-i\delta /2}%
\end{array}%
\right\} e^{-i\omega t},  \label{4column}
\end{equation}%
where $\epsilon _{1}$ and $\epsilon _{2}$ are complex amplitudes of
infinitesimal perturbations with eigenvalue $\alpha $ and wavenumber $k$.
Substituting this expression into Eq. (\ref{vnorm}) and performing the
linearization with respect to the perturbations yields four branches of the
dispersion relation:
\begin{eqnarray}
\alpha _{1,2} &=&\pm 2\sqrt{C(1-\cos k)\left[ A^{2}-C(1-\cos k)\right] }
\notag \\
&\equiv &\pm 2\sqrt{\chi },  \label{alpha12} \\
\alpha _{3,4} &=&\pm 2\sqrt{\chi +\Gamma \left[ -\Gamma +A^{2}-2C(1-\cos k)%
\right] }.  \label{alpha34}
\end{eqnarray}%
The CW\ solution is stable if condition \textrm{Re}$\left\{ \alpha \right\}
\leq 0$ holds for all four eigenvalues at all $k$. For $C\neq 0$ it is
sufficient to analyze this condition for $\alpha _{3,4}$, as inequality
\textrm{Re}$\left\{ \alpha _{1,2}\right\} \leq 0$ follows from \textrm{Re}$%
\left\{ \alpha _{3,4}\right\} \leq 0$. The subsequent consideration
demonstrates that the CW solution might be stable only if condition
\begin{equation}
A^{2}\leq C(1-\cos k)  \label{stable}
\end{equation}%
holds for all $k$, which is impossible at $A\neq 0$. Thus, for $C\neq 0$,
all the CW solutions (\ref{CW}) are modulationally unstable. This result
also suggests that dark solitons are unstable in the dimer chain, because
the corresponding CW background cannot be stable.

However, in the case of periodic boundary conditions, rather than the
infinite chain, the CW state can be stable for $C\neq 0$ if the ring-shaped
chain is sufficiently short. As follows from Eq. (\ref{stable}), unstable
are the modulational perturbations with wavelengths
\begin{equation}
\lambda \equiv \frac{2\pi }{k}>\frac{2\pi }{\arccos {\left( 1-A^{2}/C\right)
}}.  \notag
\end{equation}%
Thus, if the number of dimers in the chain with periodic boundary conditions
satisfies condition
\begin{equation}
N\leq \frac{2\pi }{\arccos {\left( 1-A^{2}/C\right) }},  \label{stab3}
\end{equation}%
which makes the length of the chain smaller than the wavelength of the
shortest unstable perturbation,the CW solution (\ref{CW}) is \emph{stable}.
Because $N$ cannot be smaller than $2$, a constraint
\begin{equation}
A^{2}\leq 2C  \label{stab4}
\end{equation}%
follows from Eq. (\ref{stab3}).

The stability analysis for the AC solution Eq. (\ref{2column}) can be
performed by substituting Eq. (\ref{4column}) into Eq. (\ref{vnorm}) with $%
C=0$. In the case, Eqs. (\ref{alpha12}) and (\ref{alpha34}) reduce to $%
\alpha _{1,2}=0$ and $\alpha _{3,4}=\pm 2\sqrt{\Gamma \left( -\Gamma
+A^{2}\right) }$, from where it follows that the AC solution is stable for
\begin{equation}
A^{2}\leq \left( A_{\max }^{(0)}\right) ^{2}\equiv \Gamma .  \label{stab2}
\end{equation}%
These solutions, stable in the AC limit, are used below to construct
numerical solutions for narrow solitons.

\section{Numerical results}

\label{NumericalResults}

Discrete evolution equations (\ref{vnorm}) were integrated numerically,
using a modification of an implicit, unconditionally stable Crank-Nicholson
scheme with accuracy $\sim \tau ^{4}$, where the time step was taken as $%
\tau =2\times 10^{-5}$. Zero boundary conditions were employed except for
Sec. \ref{UnstableDyn}, where the periodic boundary conditions were used.

\subsection{Unstable CW dynamics in the chain with periodic boundary
conditions}

\label{UnstableDyn}

Interesting dynamics was observed in the chain with periodic boundary
conditions for the initial conditions taken as per the CW solution (\ref%
{CW}), in the case when the chain's length, $N$, is only slightly
greater than the critical value specified by Eq. (\ref{stab3}),
while conditions (\ref{stab4}) and (\ref{stab2}) are satisfied.

\begin{figure}[tbp]
\includegraphics[width=\columnwidth]{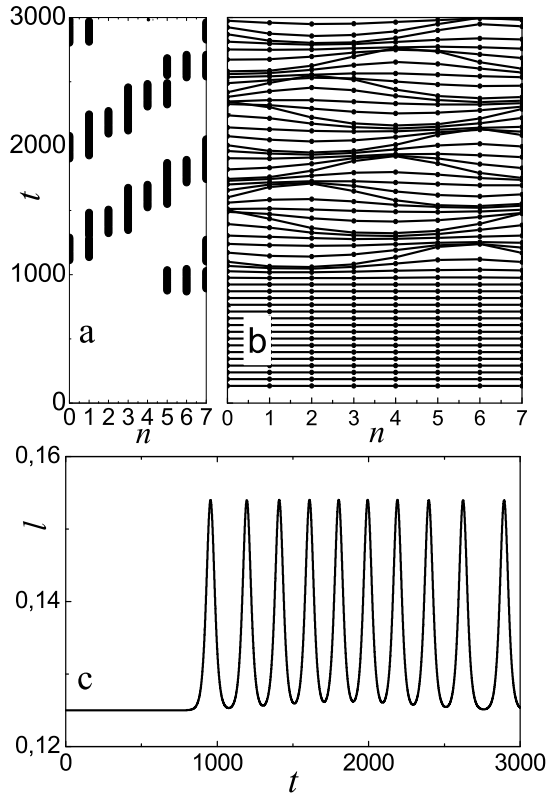}
\caption{Unstable dynamics in a compact ($N=8$) ring-shaped chain of dimers
with initial conditions corresponding to the CW state (\protect\ref%
{CW}) with amplitude $A=0.3$ and model's parameters $\protect\gamma %
=0.5$, $C=0.3$. (a) Sites (dimers) in the $(n,t)$ plane with intensity
greater than $1.1A^{2}$ are shown in black, while others are not shown. (b)
The evolution of intensity $|u_{n}(t)|^{2}$ [the picture for $\left\vert
v_{n}(t)\right\vert ^{2}$ is identical to this one, within the numerical
accuracy]. (c) Evolution of the localization parameter, $l$, defined as per
Eq. (\protect\ref{loc}).}
\label{Fig2}
\end{figure}

The instability of the system results in periodic appearance and
disappearance of weakly localized states. This dynamics is illustrated by
Fig. \ref{Fig2} for $\gamma =0.5$, $A=0.3$, $C=0.3$, $N=8$ (the
corresponding CW solution is stable for $N\leq 7$). Panel (c) in Fig. \ref%
{Fig2} displays the time variation of the localization parameter defined as
\begin{equation}
l=\left( \sum\limits_{n=0}^{N-1}{|u_{n}|}\right) ^{-2}\left(
\sum\limits_{n=0}^{N-1}{|u_{n}|^{2}}\right) \,.  \label{loc}
\end{equation}%
Note that when $\left\vert u_{n}\right\vert =\mathrm{const}$, as in the
unperturbed CW state, Eq. (\ref{loc}) yields $l=1/N$, which is the minimal
possible value, which tends to zero with the increase of $N$. On the other
hand, when single $\left\vert u_{n}\right\vert $ is different from zero, the
localization parameter attains its maximal value, $l=1$.

In these simulations, the only source of perturbations destabilizing
the CW state (\ref{CW}) was rounding errors of the numerical code.
In this case, the deviation of the solution from the initial
unstable configuration becomes visible at $t\approx 900$. The
unstable dynamics at $t>900$ results in quasi-periodic localizations
and delocalizations of $|u_{n}|^{2}$, as seen in Fig. \ref{Fig2}(c).
The maximum of $|u_{n}|^{2}$ gradually moves along the chain with a
nearly constant velocity, as seen in panels (a) and (b).

A similar effect of the periodic in time and space localization of the
unstable CW solution, subject to periodic boundary conditions, was observed
in the model of a diatomic chain of atoms \cite{we1}. A simple analytical
model describing this effects was proposed in Ref. \cite{we2}.

\subsection{Strongly localized on-site solitons}

\label{SharpSolitons}

Sharp (tightly localized) on-site solitons were constructed starting from
the AC-limit solution (\ref{2column}), which is valid for $C=0$. After
setting the initial conditions, the coupling constant was increased linearly
in time from $C=0$ at $t=0$ to $C=0.15$ at $t=400$, and then kept constant.
The total duration of the numerical run was $t=1000$. With such a slow
increase of $C$, nearly stationary solitons were readily created (details of
the stationary shape of the solitons are given below). The initial amplitude
of the soliton in the AC limit was $A=0.6$. This amplitude satisfies the AC
stability condition (\ref{stab2}) for three considered values of the
gain/loss parameter, $\gamma =0,\,0.3,\,0.9$. The chain was composed of $21$
dimers, which was sufficient for constructing tightly localized solitons.

\begin{figure}[tbp]
\includegraphics[width=\columnwidth]{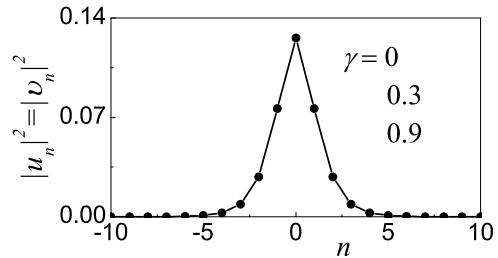}
\caption{Soliton profiles obtained from the anti-continuum limit by the slow
increase of coupling constant $C$ at three different values of the gain-loss
parameter, $\protect\gamma =0,\ 0.3,\ 0.9$. Intensities $|u_{n}|^{2}\ $and$%
\,\,|v_{n}|^{2}$ do not depend on $\protect\gamma $, while the soliton's
frequency and phase shift between $u_{n}$ and $v_{n}$ depend on $\protect%
\gamma $, as explained in the text. } \label{Fig3}
\end{figure}

Results of the simulations can be summarized as follows:

(i) The loss-gain coefficient $\gamma $ does not affect the profile and
amplitude of the soliton, $|u_{n}|^{2},\,\,|v_{n}|^{2}$. This can be seen in
Fig. \ref{Fig3}, where the shapes of the solitons constructed for $\gamma
=0,\,0.3,\,0.9$ practically overlap. Coefficient $\gamma $ only affects the
phase difference between complex quantities $u_{n}$ and $v_{n}$ and the
frequency of the soliton.

The fact that the profiles of stationary solitons do not depend on $\gamma $
can be explained by noting that if $y_{n}=W_{n}e^{-i\omega t}$ is a
stationary solution of the standard discrete nonlinear Schr\"{o}dinger
(DNLS)\ equation,
\begin{equation}
\frac{dy_{n}}{dt}=-iy_{n}+i|y_{n}|^{2}y_{n}+iC(y_{n+1}+y_{n-1}-2y_{n}),
\label{yn}
\end{equation}%
then a stationary solution of Eq. (\ref{vnorm}) can be written as
\begin{equation}
\left\{
\begin{array}{c}
u_{n} \\
v_{n}%
\end{array}%
\right\} =W_{n}e^{-i\omega t}e^{it\left( 1-\sqrt{1-\gamma ^{2}}\right)
}\left\{
\begin{array}{c}
e^{i\delta /2} \\
e^{-i\delta /2}%
\end{array}%
\right\} ,  \label{yn1}
\end{equation}%
where $\delta $ is defined by Eq. (\ref{A}). It is clear that the solution (%
\ref{yn1}) with $\gamma \neq 0$ differs from a stationary solution to Eq. (%
\ref{yn}), with $\gamma =0$, only by the frequency and phase shift, while
their intensities are same.
%\textbf{[IN THE EQUATION \ FOR }$\frac{dy_{n}}{dt}$%
%\textbf{\ I HAVE ADDED MINUS IN FRONT OF TERM }$iy_{n}$\textbf{, BECAUSE IN
%THE ANSATZ FOR }$\left\{
%\begin{array}{c}
%u_{n} \\
%v_{n}%
%\end{array}%
%\right\} $\textbf{\ WE HAVE }$\left( \omega -1\right) ,$\textbf{\ RATHER
%THAN }$\left( \omega +1\right) $\textbf{.]}
Thus, the discrete system based on Eq. (\ref{vnorm}), although being
dissipative, supports the continuous family of solitons, being in that
respect similar to the conservative DNLS equation, and different from
generic discrete dissipative systems, such as the discrete complex
Ginzburg-Landau equation \cite{GL}.

(ii) To check that the numerically found solitons are stationary
ones, we calculated the measure of non-stationarity at site $n$,
$S_{n}\equiv d\left( |u_{n}|^{2}\right) /dt$ for three solitons
constructed from the AC limit at $\gamma =0,\,0.3,\,0.9$. We found
that the result does not depend on $\gamma $, as expected in view
of the above comment. In the considered examples, the
non-stationarity of the constructed solitons is weak ($\max \left\{ {S_{n}}%
\right\}$ does not exceed $0.0003$), and it gets weaker for the
solitons constructed with the use of $C$ increasing still slower
in time.

(iii) The tightly pinned on-site-centered soliton remains stable
if its amplitude is not too large to break the $\mathcal{PT}$
symmetry. This conclusion is justified by monitoring the long-term
dynamics of the discrete solitons constructed according to the
protocol outlined above.

\subsection{Instability of inter-site and twisted tightly localized solitons}

\label{InterSolitons}

On the contrary to the stable on-site centered discrete solitons, outlined
above, the analysis has revealed that all the inter-site-centered solitons
are unstable. We attempted to construct such solitons by starting, in the AC
limit, with two adjacent excited sites, $n=0,1$, using Eq. (\ref{2column})
with $A=0.6$ as initial conditions, and gradually increasing $C$ from $C=0$
at $t=0$ to $C=0.15$ at $t=400$, and keeping then $C$ constant.

\begin{figure}[tbp]
\includegraphics[width=\columnwidth]{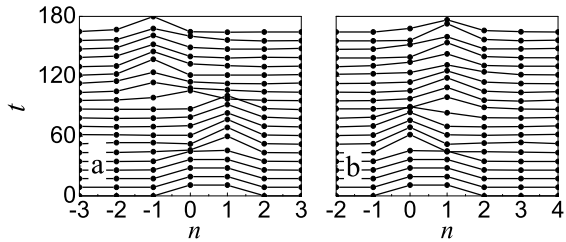}
\caption{The evolution of the inter-site-centered soliton, grown from the AC
initial conditions, following the increase of $C$ at two different values of
$\protect\gamma $. Eventually, one site spontaneously sucks the energy of
the other one. The parameters are $A=0.6$, $C=0.15$, $N=16$. In panel (a), $%
\protect\gamma =0.2$, and in panel (b), $\protect\gamma =0.6$.}
\label{Fig4}
\end{figure}

In Fig. \ref{Fig4} we present the evolution of the inter-site mode
in the course of the increase of $C$. This mode is stable only in
the absence of the coupling, $C=0$. At $C\neq 0$, one site
spontaneously sucks the energy from the other, which end us with
the establishment of a stable
on-site-centered soliton. Note that in the examples presented in Fig. \ref%
{Fig4} this happens at $t\approx 50$, i.e., before $C$ reaches the
target constant value of $0.15$. We thus conclude that the
inter-site soliton is unstable. This result seems very plausible
in view of the above-mentioned relation of stationary solitons in
this system to those in the conventional discrete NLS equation,
Eq. (\ref{yn}), where all the inter-site-centered solitons are
unstable too \cite{Panos}.

We also attempted to excite twisted (antisymmetric) solitons, by starting,
in the AC limit, with two adjacent excited sites, at $n=0$ and $1$, using
Eq. (\ref{2column}) with $A=0.6$ and introducing the phase shift of $\pi $
between the sites. This mode too turns out to be unstable at $C>0$. In the
usual lattices of the DNLS\ type, twisted solitons usually exist at large
amplitudes \cite{Panos}, but the soliton's amplitude in the present $%
\mathcal{PT}$-symmetric chain is limited by condition (\ref{stab2}), which
is a possible reason of the failure in looking for stable twisted solitons.

\subsection{Instability due to the $\mathcal{PT}$-symmetry breaking}

\label{PTinstability}

\begin{figure}[tbp]
\includegraphics[width=\columnwidth]{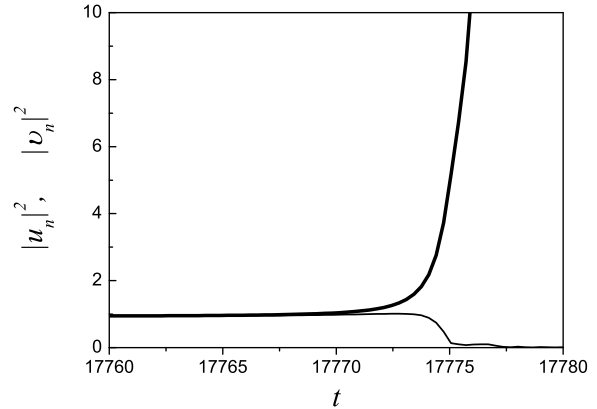}
\caption{Intensities of the fields at the central dimer, $n=0$, as functions
of time, for the soliton in the chain of dimers with a small gain-loss
mismatch, $\protect\epsilon =10^{-5}$, and $\protect\gamma =0.4$, $C=0.15$.
Thick and thin lines pertain, respectively, to the elements with gain and
loss, i.e., $|u_{n}|^{2}$ and $|v_{n}|^{2}$. The abrupt growth of $%
|u_{n}|^{2}$ signalizes the onset of the instability.}
\label{Fig5}
\end{figure}

According to the stability condition for the soliton in the AC limit, given
by Eq. (\ref{stab2}), increase in $\gamma $ results in a decrease of the
critical value of the amplitude, $A_{\mathrm{max}}^{(0)}$, above which the
localized mode becomes unstable. For the solitons in the chain with $C>0$,
the corresponding critical value was determined numerically. To this end, we
took, as the initial condition, a soliton constructed as described above
(using the slow increase of coupling constant $C$), and then switched on a
weak net pump of the energy into the soliton, by introducing a small
mismatch, $\epsilon $, in the gain/loss balance. In this case, Eq. (\ref%
{vnorm}) assumes the following form:
\begin{eqnarray}
i\frac{du_{n}}{dt} &=&+(\gamma +\epsilon )u_{n}+i|u_{n}|^{2}u_{n}+  \notag \\
&&iC(u_{n+1}+u_{n-1}-2u_{n})+iv_{n},  \notag \\
i\frac{dv_{n}}{dt} &=&-(\gamma -\epsilon )v_{n}+i|v_{n}|^{2}v_{n}+  \notag \\
&&iC(v_{n+1}+v_{n-1}-2v_{n})+iu_{n}.
\end{eqnarray}%
In the stable regime, the field intensities in all dimers remain
equal for the active and passive elements, both slowly increasing
due to small excess gain, $\epsilon >0$. The system becomes
unstable when the soliton's amplitude attains some critical value,
$A_{\mathrm{max}}$. An example of the unstable behavior of the
central dimer ($n=0$) is shown in Fig. \ref{Fig5}. In the unstable
regime, the intensity in the active element sharply increases,
while in its passive counterpart the intensity decays.

\begin{figure}[tbp]
\includegraphics[width=\columnwidth]{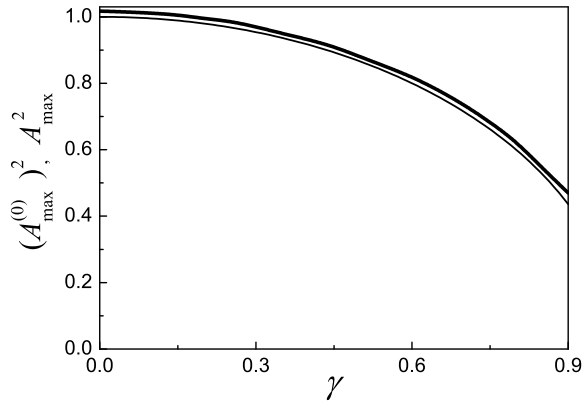}
\caption{The critical amplitude, $A_{\mathrm{max}}^{2}$, above which the
soliton's instability sets in, as a function of the loss-gain strength, $%
\protect\gamma $. The thick line is the numerical result for the solitons at
$C=0.15$. The thin line represents the critical amplitude, $\left( A_{%
\mathrm{max}}^{(0)}\right) ^{2}=\Gamma $, as predicted by the stability
condition (\protect\ref{stab2}) for the soliton in the anti-continuum limit (%
$C=0$). } \label{Fig6}
\end{figure}

In Fig. \ref{Fig6} we compare the dependence of $A_{\mathrm{max}}^{2}$ on $%
\gamma $, found numerically in this way, to the analytical result for the
local mode in the AC limit, $(A_{\mathrm{max}}^{(0)})^{2}$, as defined by
Eq. (\ref{stab2}). It is seen that they are in a good agreement, despite the
fact that $A_{\mathrm{max}}^{2}$ pertains to the chain with $C\neq 0$. The
numerical value is somewhat higher than the analytical estimate, which may
be explained by a particular criterion of instability adopted in the
numerical study: The soliton was assumed to become unstable when the largest
(over $n$) relative difference between $|u_{n}|^{2}$ and $|v_{n}|^{2}$
attained the level of $\max_{n}\left\{ ||u_{n}|^{2}-|v_{n}|^{2}|/\left(
|u_{n}|^{2}+|v_{n}|^{2}\right) \right\} >0.01$.

%We have checked weather the instability is related to the small
%excess gain, $\epsilon >0$, or it is the consequence of the
%$\mathcal{PT}$-symmetry breaking due to violation of the condition
%Eq. (\ref{stab2}), limiting the soliton amplitude. To do so, we
%set $\epsilon =0$ after the soliton amplitude reached the value
%$1.001(A_{\mathrm{max}}^{(0)})^{2}$. It was found that even at
%$\epsilon =0$ the soliton with the amplitude slightly exceeding
%the limiting value was destroyed due to the instability.

Results presented in Figs. \ref{Fig5} and Fig. \ref{Fig6} were
obtained for the inter-dimer coupling constant $C=0.15$. Similar
results were obtained for other values, up to $C=0.5$. In all the
cases considered, the numerically found instability threshold,
$A_{\mathrm{max}}^{2}$, was very close to
$(A_{\mathrm{max}}^{(0)})^{2}$ predicted by Eq. (\ref{stab2}) for
the AC-limit soliton. Thus, the threshold practically does not
depend on the coupling constant $C$. This result demonstrates a
drastic difference in the
dynamical properties of the discrete solitons in the $\mathcal{PT}$%
-symmetric chain from its counterparts in the usual DNLS equation, where the
instability of various soliton modes is solely determined by $C$ \cite{Panos}%
.

\begin{figure}[tbp]
\includegraphics[width=\columnwidth]{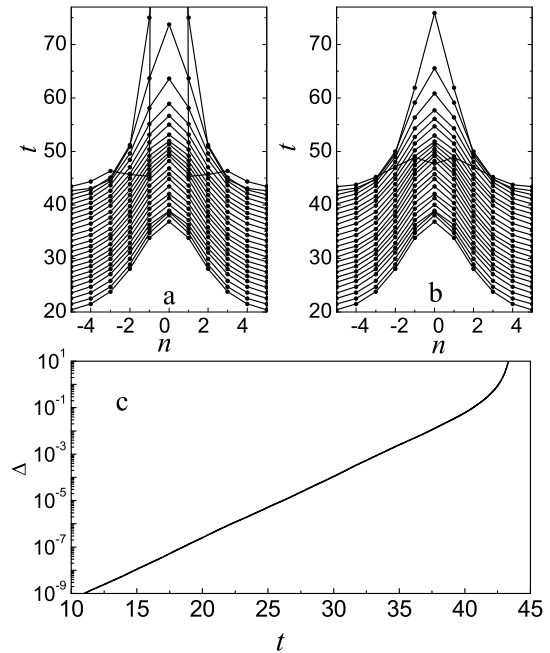}
\caption{The soliton dynamics in the unstable regime. Panels (a) and (b)
display $|u_{n}(t)|^{2}$ and $|v_{n}(t)|^{2}$, respectively. Panel (c) shows
the difference in the intensities between the active and passive elements at
the central dimer ($n=0$), $\Delta =||u_{0}|^{2}-|v_{0}|^{2}|$, as a
function of time. Parameters are $\protect\gamma =0.9$, $C=2.3$, $A=0.9$, $%
V=0$.} \label{Fig7}
\end{figure}

The dynamical blowup of unstable solitons is displayed in Fig.
\ref{Fig7}.
Initial conditions were set using approximation (\ref{discsol}) with $%
x_{0}=0 $. In this case, $A^{2}=0.81>\Gamma =\sqrt{1-\gamma
^{2}}\approx 0.436$, hence the AC stability condition Eq.
(\ref{stab2}) is violated. The result displayed in panel
\ref{Fig7}(c) suggests that, for $t<40$, the difference $\Delta
\equiv \left\vert |u_{0}(t)|^{2}-|v_{0}(t)|^{2}\right\vert $
increases with time exponentially, and then it starts to grow
still faster. This change in the unstable dynamics is due to a
qualitative change in the soliton's profile,
that can be seen in panels \ref{Fig7}(a,b). At $t<40$, intensities $%
|u_{0}|^{2}$ and $|v_{0}|^{2}$ grow with time, but, after the difference
between them becomes sufficiently large, $|u_{0}|^{2}$ starts to grow
faster, while $|v_{0}|^{2}$ begins to decrease.

Results presented in Fig. \ref{Fig7} were obtained for the
quiescent soliton
($V=0$). We have also studied the influence of the soliton's velocity, $%
0<V\leq 0.5$, on the onset and development of the instability. It was found
that the growth rate of $\Delta $ in the unstable regime and the critical
soliton amplitude $A_{\mathrm{max}}^{2}$ do not depend on $V$.

This instability can be explained by comparison with the system of
two linearly-coupled DNLS\ equations, where the symmetric discrete
solitons is destabilized by the symmetry-breaking bifurcation
\cite{discrete}. It seems plausible that the spontaneous trend to
the symmetry breaking, due to the self-attraction in each chain,
explains the onset of the instability in the present setting.
However, a drastic difference from the dual-core DNLS system,
where the symmetry breaking replaces the original destabilized
soliton by stable asymmetric ones, in the $\mathcal{PT}$ system
this is impossible, as asymmetric modes cannot maintain the
balance between the gain and loss, and instead exhibit the blowup,
as shown in Fig. \ref{Fig7}.

\subsection{The Peierls-Nabarro potential}

\label{PNP}

\begin{figure}[tbp]
\includegraphics[width=\columnwidth]{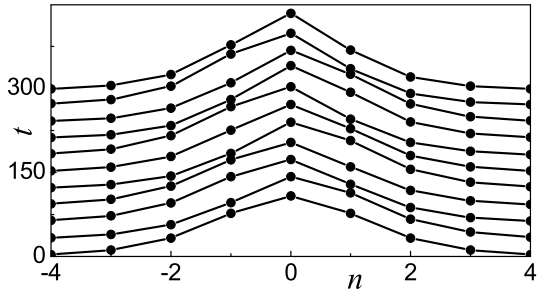}
\caption{The effect of the Peierls-Nabarro (PN)\ potential on the soliton
dynamics. The initial momentum imparted to the soliton is not sufficient to
overcome the PN potential barrier, therefore the soliton oscillates near a
local minimum of the potential. The parameters are $A=0.6$, $V=0.02$, $%
\protect\gamma =0.5$, $C=0.2$.} \label{Fig8}
\end{figure}

The effect of the Peierls-Nabarro (PN) potential in the chain of
dimers is demonstrated by simulations presented in Fig.
\ref{Fig8}, where the initial conditions were set by using Eq.
(\ref{discsol}) with $x_{0}=0$. The parameters were chosen so as
to produce a relatively sharp soliton when the PN barrier is
noticeable. The soliton was kicked with a relatively small
velocity ($V=0.02$). The momentum given to the soliton was not
sufficient to overcome the PN barrier, therefore, instead of
propagating along the chain, the soliton oscillates in a local
well of the potential.

\subsection{Collisions between moving solitons}

\label{Collisions}

\begin{figure}[tbp]
\includegraphics[width=\columnwidth]{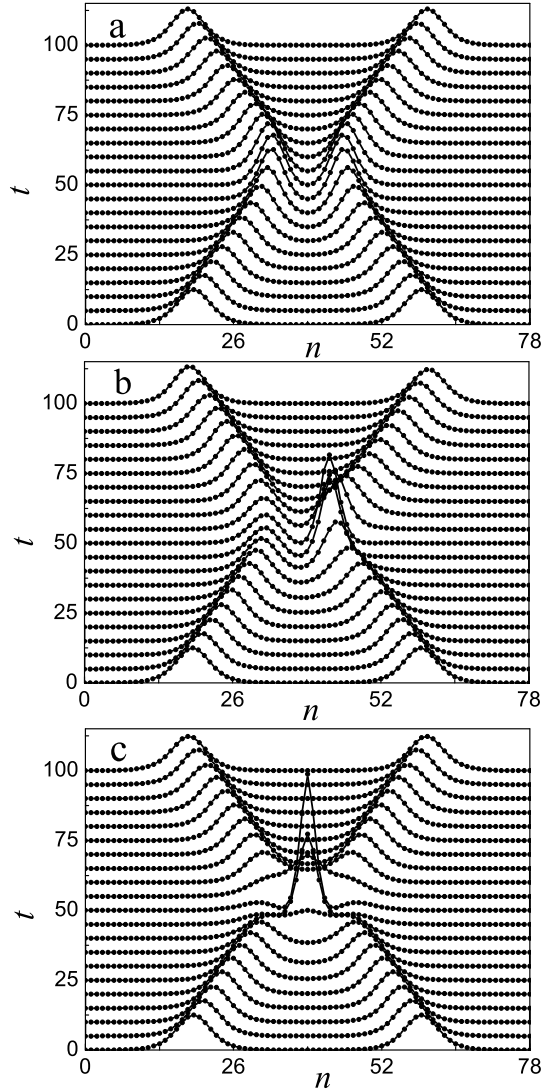}
\caption{Examples of quasi-elastic collisions between solitons for $\protect%
\gamma =0.5$, $C=2$, $A=0.5$, and velocities $V=\pm 0.3$. The phase shift
between the solitons in panels (a), (b), and (c) is, respectively, $\protect%
\pi $, $\protect\pi /2$, and $0$. } \label{Fig9}
\end{figure}

The existence of moving solitons suggests a possibility to study
collisions between them. Examples of collisions between identical
solitons with opposite velocities ($V=\pm 0.3$) are presented in
Fig. \ref{Fig9}. In this
case, the solitons are relatively wide, hence the approximate solution (\ref%
{discsol}) can be used for setting the initial conditions. The collisions
are seen to be nearly elastic, with the solitons almost completely
recovering their original shapes.

The solitons with phase shift $\pi $ in (a) repel each other, while the
in-phase ones in (c) interact attractively. The collisions of the repelling
out-of-phase solitons does not result in an overlap of their cores. For the
phase shifts different from $\pi $, the overlap takes place, hence the
largest intensity at the collision point is greater than in isolated
solitons [see panel (b)], attaining the maximum for the in-phase collisions,
as can be seen in (c). It is worthy to note that, although the instantaneous
value of the amplitude at the collision point may be considerably larger
than the critical amplitude defined by Eq. (\ref{stab2}), this does not
result in the onset of the instability, as the collision time is
insufficient for that.

\section{Conclusions}
\label{Conclusions}

We have introduced a novel model describing an array of harmonically coupled $\mathcal{%
PT}$-symmetric dimers. Each dimer consists of linearly coupled active and
passive elements with balanced gain and loss. The nonlinearity is represented
by the conservative cubic term in the equation of motion for each element.

Using numerical simulations and analytical approximations, it was
demonstrated that, in the regime of the sufficiently strong inter-chain
coupling, the weakly discrete chain supports the free motion of solitary
waves. At sufficiently small velocities, the moving solitary waves do not
radiate energy and collide practically elastically, while solitons moving
with large velocities emit radiation.

The chain of dimers was demonstrated to support tightly localized (sharp)
stationary solitons in the regime of weak inter-dimer coupling (strong
discreteness). Naturally, these strongly pinned solitons are immobile. A
relation between the stationary solutions in the chain of dimers and the
solutions to the standard discrete nonlinear Schr\"{o}dinger
equation was established.

Both wide and sharp solitons develop an instability if their amplitudes
exceed the critical value, which leads to the blowup. This is a specific
feature of the $\mathcal{PT}$-symmetric chain, making the dynamical
properties of the discrete solitons different from those in the usual DNLS
system, in spite of the similarity in their shapes. An explanation to this
feature, based on the trend to the spontaneous symmetry breaking, induced by
the intrinsic self-attraction in the parallel chains, was proposed. The
critical value decreases with the increase of the gain-loss parameter. Above
the instability threshold, the $\mathcal{PT}$-symmetry breaking occurs,
resulting in the divergence of the intensity at the active element, and
decay of the intensity at the lossy one. A noteworthy finding is the
independence of the critical amplitude on the lattice coupling constant and
soliton's velocity.

To continue the work in this direction, it would be interesting to study the
chains consisting of more complex $\mathcal{PT}$-symmetric elements (such as
those with the nonlinearity of the gain and loss \cite{nonlin2}), and to
extend the analysis for two-dimensional lattices.

\section*{Acknowledgments}

B.A.M. and S.V.D. appreciate hospitality of the Nonlinear Physics Center at the
Australian National University. The work was partially supported by the
Australian Research Council.

\end{document}